\documentclass[12pt,preprint]{aastex}

\begin{document}

\title{{\it Spitzer} observations of Bow Shocks and Outflows in RCW 38. }

\shorttitle{Bow Shocks \& Outflows in RCW 38}

\author{E. Winston\altaffilmark{1}, S. J. Wolk\altaffilmark{2} , T. L. Bourke\altaffilmark{2}, S.T. Megeath\altaffilmark{3}, R. Gutermuth\altaffilmark{4,5}, B. Spitzbart\altaffilmark{2}}

\altaffiltext{1}{ESA-ESTEC (SRE-SA), Keplerlaan 1, 2201 AZ Noordwijk ZH, The Netherlands. }
\email{ewinston@rssd.esa.int}
\altaffiltext{2}{Harvard Smithsonian Center for Astrophysics, 60 Garden St., Cambridge MA 02138, USA.} 
\altaffiltext{3}{Ritter Observatory, Dept. of Physics and Astronomy, University of Toledo, 2801 W. Bancroft Ave., Toledo, OH 43606, USA. }
\altaffiltext{4}{Five Colleges Astronomy Department, Smith College, Northampton, MA  01027}
\altaffiltext{5}{Department of Astronomy, University of Massachusetts, Amherst, MA  01003}

\begin{abstract}

We report {\it Spitzer} observations of five newly identified bow shocks in the massive star-forming region RCW 38.  
Four are visible at IRAC wavelengths, the fifth is only visible at 24~${\mu}m$.
{\it Chandra} X-ray emission indicates that winds from the central O5.5 binary, IRS~2, have caused an outflow to 
the NE and SW of the central subcluster.  
The southern lobe of hot ionised gas is detected in X-rays; shocked gas and heated dust  from the 
shock-front are detected with {\it Spitzer} at 4.5 and 24~${\mu}m$.  
The northern outflow may have initiated the present generation of star formation, based on the filamentary 
distribution of the protostars in the central subcluster.   
Further, the bow-shock driving star, YSO~129, is photo-evaporating a pillar of gas and dust.  
No point sources are identified within this pillar at near- to mid-IR wavelengths.  

We also report on IRAC 3.6 \& 5.8${\mu}m$ observations of the cluster DBS2003-124, NE of RCW 38,  
where 33 candidate YSOs are identified.  
One star associated with the cluster drives a parsec-scale jet.  Two HH objects associated with the jet are visible 
at IRAC and MIPS wavelengths.  The jet extends over a distance of $\sim$3~pc.  
Assuming a velocity of 100~km/s for the jet material gives an age of 3$\times$10$^4$yr, indicating that 
the star (and cluster) are likely to be very young, with a similar or possibly younger age 
than RCW~38, and that star formation is ongoing in the extended RCW~38 region.

\end{abstract}


\today

\section{\bf Introduction}

Young stars are not only shaped by their natal environments but shape it in return.  They are
responsible for triggering further generations of star formation, ablating the 
circumstellar disks of their lower mass neighbours, clearing gas and dust in parsec-scale 
cavities surrounding them, and altering the processing of dust grains in the intercluster medium.   
Observations of young clusters in the infrared (IR) and X-ray have been combined to allow the identification 
of the stellar population in these regions from the most massive down to the substellar \citep{prei}.  Such surveys indicate that perhaps more 
than 60\% of young stars form in  clusters \citep{car,meg2}.  Our own Sun is thought to have formed 
in a massive cluster and have assimilated material (such as $^{60}Fe$ and other short-live nucleotides) 
from a near-by supernova \citep{hes}.  The study of the effects 
of high mass stars in star forming regions is thus of great importance in the understanding of both the 
large scale development of clusters and of the smaller scale evolution of planetary systems around 
lower mass Sun-like stars.  Such studies indicate that the sub-structure within these sites of star formation 
is complex and this can have an impact on the natal environment of individual stars.  

Nearby regions of high mass star formation have recently been shown to exhibit varying spatial morphologies, 
such as the compact spherical structure of Tr~15 \citep{wan11} or the subclustered Tr~16 \citep{wol11}, where 
the clusters are contigous and show no sign of mass segregation.    
RCW~38 itself shows evidence for at least four subclusters which are more loosely associated than those of 
Tr~16 and show indications of mass segregation in the central subcluster but not in the surrounding subclusters.  
By studying a number of high mass clusters we can better understand how the spatial distribution of young 
stars and their exposure to photo-ionising radiation can affect the fomation and evolution of young stars and 
their planets.  

One way that massive stars interact with their environment is through strong stellar winds.  These winds can clear gas 
from their environs and in so doing generate shocks where the high velocity outflowing gas intersects 
with the ambient interstellar/cluster medium or with the winds from other local massive stars \citep{dra}.   
Termination shocks arise where the stellar wind terminates as it is balanced by the ram 
pressure from the interstellar medium (ISM) or the wind from a local massive star.  If the relative 
velocites of the star and ISM/wind are large the shocked material is swept backwards around the 
star forming a 'bow shock'. 
The orientation of the bow shock and its 'stand-off' distance, the distance from the generating 
star to shock, are determined mainly by the the velocity of the driving star and its wind, and the 
flow of the ISM or massive star wind.

Parsec scale collimated outflows (jets) from young protostars entrain surrounding gas and dust 
and can disrupt their parent clouds, perhaps triggering star formation in regions where over densities
are generated or bringing it to an end by removing the gas from which young stars form.  
Massive stars with their strong stellar winds are particularly effective in this regard, but Herbig AeBe and 
T Tauri stars are also known to drive outflows.  Such outflows are usually identified in the visible and 
near-IR as Herbig-Haro (HH) objects or knots of roughly symmetric emission around a YSO \citep{bal2,bal3}.  
X-ray emission from jets has also been observed, e.g.  in DG Tau \citep{gud} and  L1551 IRS~5 \citep{sch}.
The lifetimes of these jets are short, typically $\sim$10$^5$~yr.  They are typically most active during the high accretion protostellar 
phase, but can continue into the class II phase.  The knots of emission may be linked to episodic accretion \citep{bal2,rei}.

Recently, observations of cluster populations in the mid-IR with $Spitzer$ have identified emission from outflows, HH-objects 
and bow shocks in star forming regions.   \citet{pov} report the identification using the {\it Spitzer} GLIMPSE survey data, of six 
bow shocks in two massive star forming regions, M17 and RCW49.  
\citet{tak} have modeled emission from six HH-objects over the IRAC bands and have found that the emission is mainly due to 
H$_2$ emission and unlikely to have significant contributions from H~I, [Fe~II], fluorescent H$_2$, or PAH emission.  
The HH objects HH 54 and HH 7-11 have been the subject of a recent study by \citet{neu} with $Spitzer$ IRS that mapped 
emission from rotational lines of molecular hydrogen (associated with the IRAC bandpasses) and fine structure transitions of 
ions such as Fe$^+$ (associated with the MIPS 24~${\mu}m$ bandpass).  The authors found that the H$_2$ emission was consistent 
with nondissociative shocks with velocities of $\sim10-20~km s^{-1}$ and the fine structure emission with faster shocks 
with velocities of $\sim35-90~km s^{-1}$ offset from the slower shocks,  indicating that shocks detected at 24~${\mu}m$ may 
be faster dissociative shocks than those detected in the IRAC bandpasses only. 

Bow shocks have also been considered as indicators for run-away O-stars in young massive clusters \citep{gva}.  
The orientation of their bow shocks relative to the cluster, pointing away from its centre, implies the star has a velocity 
which is taking it out of the cluster.    Such observations indicate that the O-stars in question form centrally and their current 
trajectories arise from dynamical interactions in the cluster centre \citep{kro}.

The RCW 38 cluster is a region of high mass star formation, lying at a distance of about 1.7~kpc from Earth.   
The region is one of the closest high mass star forming regions known, and contains an estimated $10^3$ to $10^4$ members, 
including more than thirty O star candidates  \citep{wol06,winR}. An extensive 
overview of the previous studies of RCW~38 is to be found in the Handbook of Star Formation 
article on the cluster \citep{wolkhandbook}.   The massive star candidates are spatially distributed throughout the 
observed region but are concentrated in four subclusters.   The central subcluster of RCW~38 is dominated by two 
bright mid-IR sources: the O5.5 binary IRS~2 and the bright emission ridge IRS~1 \citep{fro,smi,der}. 
\citet{smi} suggest that the central region is a cavity blown by the wind from the IRS~2 binary and 
that IRS~1 is a ridge of material swept into a shell around IRS~2.  

In our previous paper on the RCW~38 region, \citet{winR}[hereafter referred to as $Paper~1$], 
we presented mid-IR observations taken with $Spitzer$ IRAC \& MIPS combined with $Chandra$ 
archival data.  
In this paper we will report on the identification of five bow shocks surrounding YSOs 
in the RCW~38 star forming region.  We will then discuss the larger scale evolution of 
the cluster due to the winds and/or novae of massive stars in the central subcluster.  
Lastly we will discuss the embedded cluster to the NE of RCW~38: DBS2003-124, first 
identified in a survey for new IR clusters in 2MASS by \citet{dut}, and a newly identified 
protostellar jet associated with a young star in that cluster.

\section{\bf Observations and Data Reduction}\label{obsdata} 

The observations and data reduction for the $Spitzer$ and $Chandra$  data were presented in $Paper~1$ and \citet{wol06}.  
A brief summary of that work is presented here.   We obtained observations of the RCW~38 
region with the {\it Spitzer} InfraRed Array Camera (IRAC; \citet{faz}) in four wavelength bands: 3.6, 4.5, 5.8, 
and 8.0~$\mu$m and with the Multiband Imaging Photometer for {\it Spitzer} (MIPS) \citep{rie}. 
This photometry was supplemented by $J$, $H$ and $K$-band photometry from the 2MASS point source 
catalogue \citep{skr}.  The infrared observations were further combined with {\it Chandra} Advanced CCD Imaging 
Spectrometer (ACIS; \citet{wei})  X-ray data.   

In $Paper~1$ we identified 624 young stellar objects in RCW 38, of which, 226 were detected in X-rays with {\it Chandra}.  
The YSOs were of the following evolutionary classes \citep{lad,win2007}:  23 class I, 90 flat spectrum, 437 class II, and 74  class III.  
Of the YSOs, 9 class I, 18 flat spectrum, 125 class II, and 74 class III were also detected in the {\it Chandra} X-ray 
observation.  The spatial distribution was studied using YSO density maps and minimum spanning-tree techniques 
and found to show evidence of three subclusters surrounding the central IRS2 
subcluster.   One of these subclusters is Obj-36 \citep{dut} associated with the reflection nebula vdBH-RN43 \citep{vand} 
and has now been identified as a subcluster of the RCW~38 complex.      
We found that the observed relationship of $N_H$/$A_K$ was consistent with the \citet{vuo} ratio of $1.6 \times 10^{22}$ 
for the local ISM and nearby molecular clouds.
An examination of photometric variance identified 72 variable sources, with variables in each of the subclusters 
and in the dispersed population. Using near-IR color-magnitude criteria, 29 candidate O stars were identified.  These were also 
distributed across the extended RCW~38 region, with massive stars associated with each of the subclusters.

The bow shocks identified in this paper have an extended geometry such that their fluxes could not be obtained using the point 
source extraction methods applied to the stellar sources in $Paper~1$.  The flux apertures were constructed manually  
to most accurately enclose the shock on the image.  The area of each bow shock was estimated 
using an irregular polygonal aperture in $ds9$ \citep{joy} and the flux estimated using $funtools$ \citep{man}. 
The same aperture was used at each IRAC band, being defined in the band where emission in that shock was 
strongest and best defined, and then replicated at each wavelength.  
The background sky emission was estimated and removed by taking the flux in an adjacent rectangular aperture and 
scaling to the area of the shock aperture.  The 24~${\mu}m$ bow shock fluxes were estimated using $funtools$, similarly 
to the IRAC fluxes, using the same irregular polygonal aperture used in the IRAC bands where the shock was detected.

\section{\bf Discussion}

\subsection{The Bow Shocks}

\begin{figure}
\epsscale{1.}
\plotone{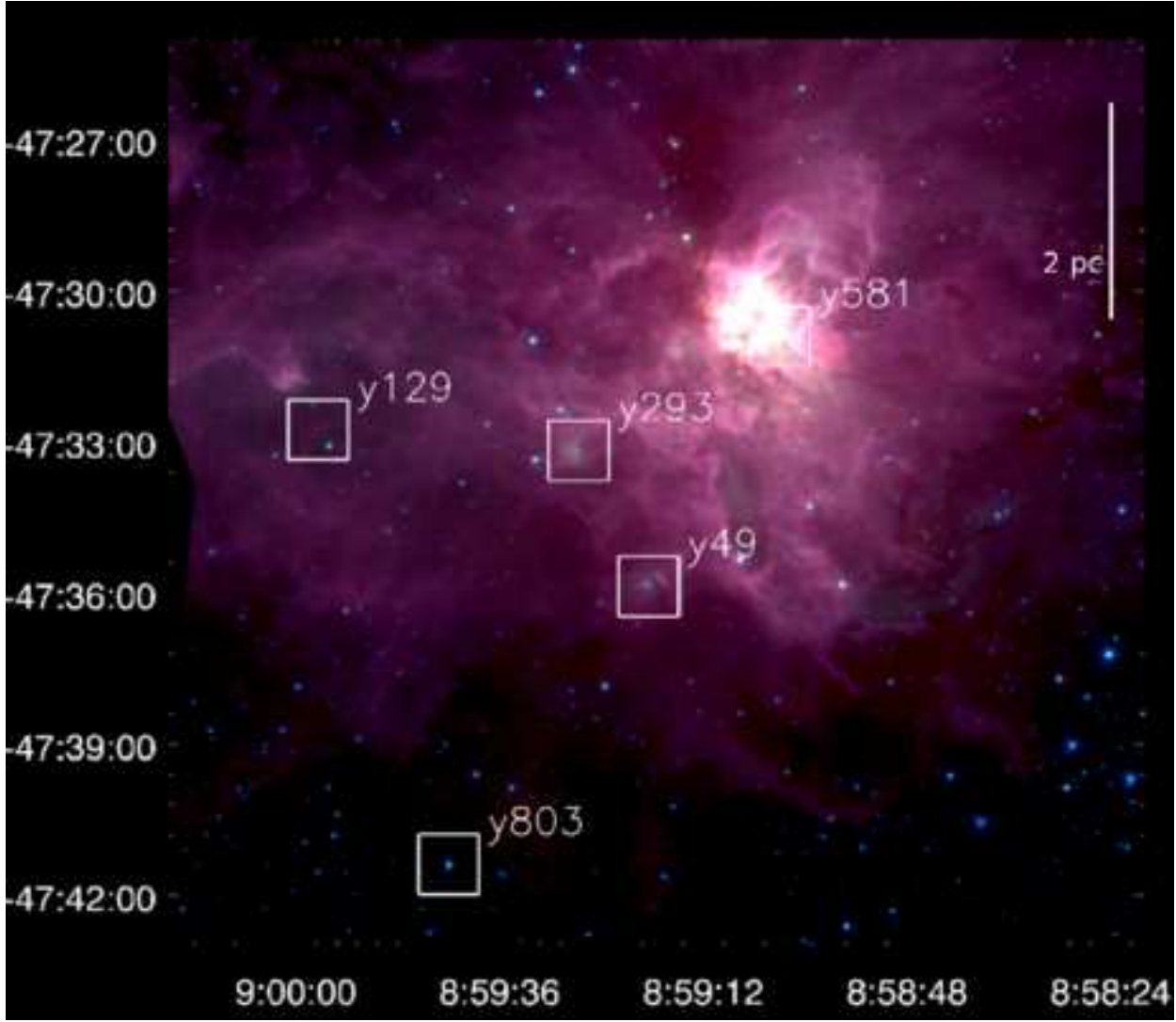}
\caption[]{ Three band false color image of the RCW 38 cluster using the IRAC bands 3.6 ${\mu}m$ in blue, 
4.5 ${\mu}m$ in green and 8.0 ${\mu}m$ in red. The positions of the five identified bow shocks are indicated. }
\label{fig1}
\end{figure}

\begin{figure}
\epsscale{.7}
\plotone{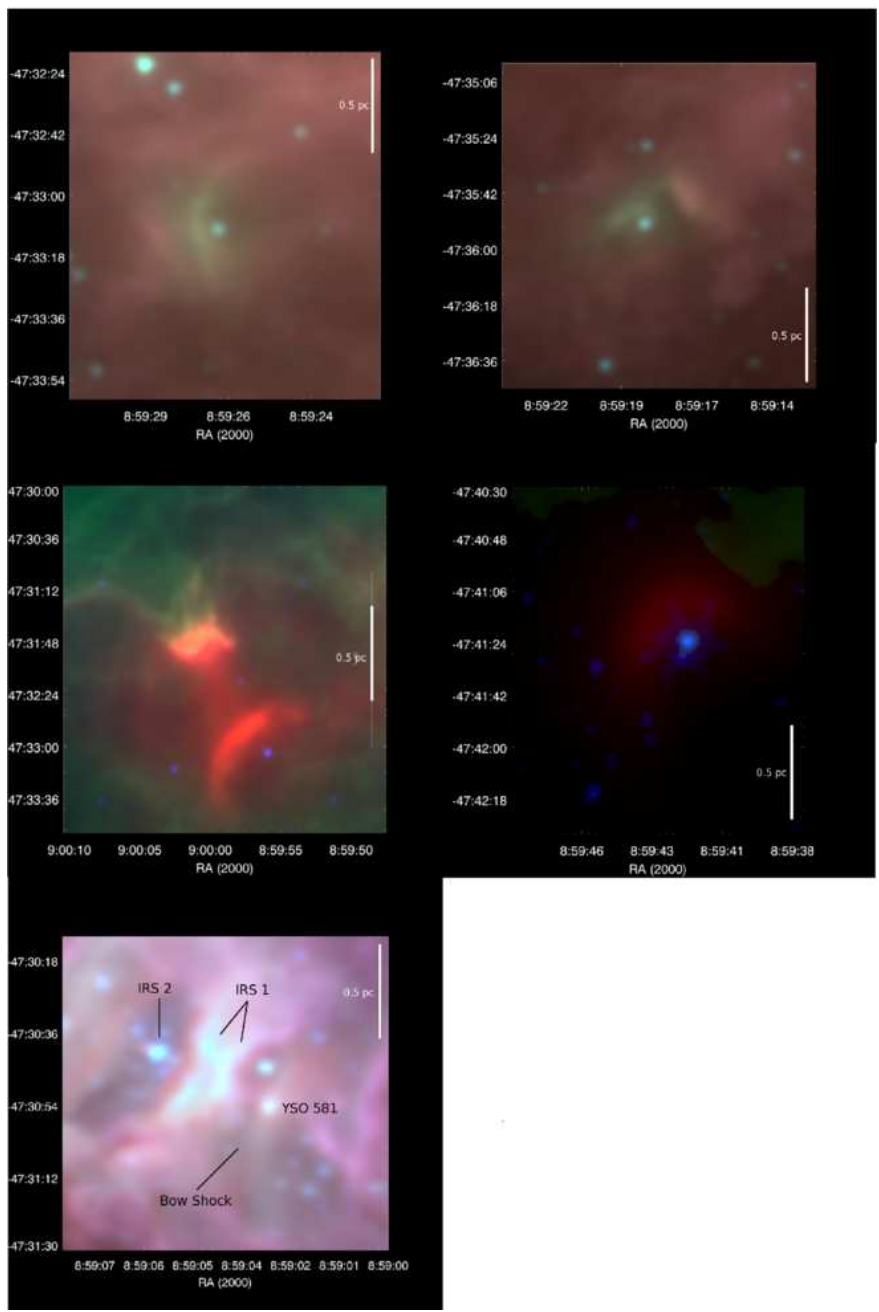}
\caption[]{
Enlarged view in three band false color IRAC/MIPS images of the five bow shocks in RCW 38.  
{\bf Above Left:} The bow shock around YSO 293 in IRAC bands 3.6 ${\mu}m$, 4.5 ${\mu}m$ and 8.0 ${\mu}m$. 
{\bf Above Right:}  The YSO 49 and bow shock in IRAC bands 3.6 ${\mu}m$, 4.5 ${\mu}m$ and 8.0 ${\mu}m$. 
{\bf Central Left:} The YSO 129 in IRAC bands 4.5 ${\mu}m$ and 8.0 ${\mu}m$ and MIPS 24${\mu}m$, with the 
bow shock only faintly visible in bands other than the 24 ${\mu}m$. The evaporating pillars are also bright at 8.0 and 24 ${\mu}m$.  
{\bf Central Right:} The fainter shock from YSO 803, to the SE in IRAC 4.5 ${\mu}m$ and 8.0 ${\mu}m$ and MIPS 24 ${\mu}m$. 
It is only visible at 24${\mu}m$. 
{\bf Below:} The central shock from YSO 581 is shown in a  three color image of IRAC 3.6 ${\mu}m$, 4.5 ${\mu}m$, 8.0${\mu}m$. 
IRS1 glows brightly to the E of the bow shock, which is likely interacting with the strong winds from IRS2 further to the east in the image.  }
\label{fig2}
\end{figure}

In RCW~38 we identify five bow shocks in highly differing environments.  
The coordinates and photometry for the five driving stars are given in Table~\ref{tablestars}. 
Their individual identifiers correspond to the $Spitzer$ IDs listed in Table~2 of $Paper~1$, where 
their evolutionary classification and variability over our two epochs of IRAC observations were determined.   
The coordinates of the apexes of the bow shocks and their estimated fluxes are listed in Table~\ref{tableshocks}.  
Figure~\ref{fig1} presents an IRAC false-color image of the RCW~38 cluster with the locations 
of the five identified shocks highlighted. The bow shocks are clearly visible in shocked hydrogen 
emission,  seen predominantly at 4.5${\mu}m$ (in green). PAH emission from heated dust is visible at 8.0${\mu}m$.   
The region of bright yellow emission to the west of centre in the image is the central subcluster of RCW~38 
where IRS~1 \& 2 are located.  
Figure~\ref{fig2} provides IRAC and MIPS false-color enlarged images of each of the five identified shocks 
and their driving stars.  The upper two plots (YSO~293 \& YSO~49) show the bow shocks clearly visible in shocked hydrogen 
at 4.5${\mu}m$.  The shocks in the central plots (YSO~129 \& YSO~803) emit predominantly at 24${\mu}m$, indicating 
perhaps faster dissociative shocks \citep{neu}.  The lower plot shows YSO~581, which is seen as a faint emission feature  
against IRS~1, the bright X-shaped emission feature to the NE of the shock in the centre of the image.  
IRS~2 is the bright blue source to the E of IRS~1.  
Details on the individual driving stars and bow shocks can be found in the appendix.   
  
The near- and mid-IR  fluxes were used to construct spectral energy distributions (SEDs) for the five driving 
stars and bow shocks, with the plots of the SEDs given in Fig.~\ref{fig3}.  The star symbols show the stellar 
fluxes while the squares indicate the fluxes of the shocks.  The shock fluxes of YSO~803 and YSO~129 are very faint 
shortwards of 24${\mu}m$ and so the IRAC fluxes are not greatly in excess of the background and are more 
uncertain.  Both the stellar and shock fluxes from YSO~581 are contaminated by the superposition of the shock 
and star due to the line of sight.  
The objects YSO~581, YSO~293, and YSO~49 were classified as C0/I and II sources as their stellar SEDs exhibit 
excess emission above that of a stellar photosphere.   
The SEDs of the bow shocks are consistent with emission from a black body.  
These findings are consistent with those of shocks in RCW49 and M17 presented by \citet{pov}, and by \citet{bal} 
in the Orion Nebular Cluster. The ONC is relatively nearby at a distance of 145~pc, M17 lies at a similar distance 
(1.5~kpc) to RCW~38, while the distance to RCW~49 is given as 4.2 or 6~kpc in \citet{pov}.  The five RCW~38   
shocks increase the number of known mid-IR shocks by $\sim$40\% and more than double the known sample at 
1-2~kpc.

\subsubsection{Dynamical Implications}

Consideration of the stand-off distance of the bow shock (the distance from the star to the apex of the shock) 
and the mass loss rate of the driving star can lead to an estimation of the relative velocity of the stellar winds or ISM in the region.  
From \citet{pov}, following the normalisation of \citet{van}, we have an equation relating the velocity ($v_{o}$) 
and hydrogen particle density ($n_{o,3} = 10^{3} cm^{-3}$) of the ambient medium to the mass loss rate 
($\dot{M}_{w,-6} = 10^{-6} M_{\sun} yr^{-1}$) and relative velocity ($v_{w,8} = 10^{8} cm s^{-1}$) of the driving 
star and the bow shock stand-off distance ($d_{w}$)
\begin{eqnarray*}
v_{o} n_{o,3}^{1/2}  =  1.5 (\frac{d_{w}}{pc})^{-1}  (\dot{M}_{w,-6} v_{w,8})^{1/2}  
\end{eqnarray*}
The stand-off distance must incorporate a $cos(i)$ term since we cannot ascertain the angle of the shock with 
respect to the plane of the sky. However, $cos(i)$ has a mean value of 0.9 for angles between 0$^{o}$ and 
45$^{o}$ so this will affect the resulting velocity measurement at the 10\% level.   The density of the ISM is 
assumed to be approximately $10^3~cm^{-3}$, and hence equal to unity in the equation.   The mass loss rates and 
stellar wind velocities from young massive stars are highly variable and strongly dependent on stellar spectral 
type \citep{ful}.  This uncertainty is further complicated by the estimation of the spectral type, which is calculated 
from the near-IR colors and not spectroscopy.  In this case a value of $<$0.2$\dot{M}_{w,-6}$ is applied to all 
stars with spectral type later than O9 applicable to all five objects in this case.  This value is based on estimates taken 
from \citet{vin} and \citet{ful}.   A thorough description of the assumptions made in this 
calculation is presented in \citet{pov}.   

The offset in pixels from the coordinates of the star to the apex of the bow shock was measured on the 8~${\mu}m$ 
mosaic and converted to distance in parsecs  assuming a distance to RCW~38 of 1.7~kpc.  
The spectral types of the driving stars were estimated from their positions on the $K$ v. $J - K$ color-magnitude 
diagram.  The spectral types, stand-off distances and estimates of the relative velocity for the shocks are given in 
Table~\ref{tableprops}, except YSO 581, the geometry of which precludes the measurement of the stand-off 
distance.    With the caveat that the values of $v_{o} n_{o,3}^{1/2}$ are highly uncertain we can say that in 
the cases of the four shocks with measurements the velocities span a small range, from 4-12~$kms^{-1}$.  
The velocities calculated for sources 129 and 803, those only visible at longer wavelengths, are somewhat lower 
than those of the sources with shocks clearly visible in the IRAC bands.  
This would not agree with their emission arising from fast dissociative shocks and is perhaps an indication that 
these more outlying sources are located in a lower density medium.   It may also be linked to evolutionary 
classification: both are class III sources. It may also be related to their being shocked by the ISM and not due to 
interaction with wind from high mass stars.

The orientations of all five bow shocks in RCW~38 appear to be aligned with O-star candidates identified in their vicinity.  
None of the five shocks are oriented such that their driving stars would appear to be following a trajectory away from IRS~2.  
This allows us to rule out the possibility that the stars driving the bow shocks are run away O-stars that are being ejected from the central subcluster.   
YSOs 293 and 49 are aligned to oc15-v13 (see Table 11 in $Paper~1$), while the apex of the central bow shock is clearly 
pointing towards the central O5.5 binary IRS2.  This orientation indicates that the shock is $\sim60\deg$ SW 
of the binary and also perhaps $\sim15-45\deg$ behind the O stars.  This would imply that the O star binary is 
not clearing out a cylindrical structure as this is a projection effect and these walls lie behind the massive stars.  
Their winds are likely compressing and illuminating the posterior wall of the H$_2$ blister bubble.  
In the case of YSO 129 and YSO~803 the source of the interaction is less clear; it may be the candidate O-star NE20, 
associated with the new subcluster, c.f. Sec.~\ref{necj}. 
However it is more likely that they trace the direction of the flow of the ISM material in their vicinity, implying that this 
material is moving in a roughly E to W direction across the face of the cluster.

\subsection{Outflows \& Winds: Influence of the massive stars}\label{ow}

Discussion of the central O stars of IRS~2 leads to an examination of the overal structure of 
RCW~38 and the effect of the massive stars and their winds on the evolution of the cluster. 
In planetary nebulae, complex X-ray emission structures have been observed which show 
both evidence of axial symmetry and an anticorrelation between the X-ray diffuse emission 
and the infrared extinction, indicating that the X-ray outflows can in part be shaped by the 
material into which they expand \citep[cf:][]{kas,chu}.  Similar features have also been observed 
in the centre of RCW~38.   
The two O5.5 stars of IRS~2 have an estimated combined luminosity $log(L)$ of 5.71~$L_{\sun}$ and 
hydrogen ionising photon flux $log(Q_{H})$ of 49.4  \citep{der,smith,mar} and are considered 
likely to be the main drivers of the outflows from the central subcluster.  \citet{wol02} modelled the 
diffuse X-ray emission and found it was best fit by synchotron emission.  
We infer from the orientation of the central bow shock of YSO~581, that the central 
O-stars of IRS~2 are not hemmed in by the dense dust and gas structures of IRS~1 and 
surrounding rim of material. Rather, they may be some distance in front of the face of the molecular cloud,  
perhaps $\sim$0.3~pc, (the radius of a sphere centred on IRS~2 and defined by the rim of emission at 8~${\mu}m$ 
at the assumed distance to RCW~38 of 1.7kpc).  
The outflows and stellar winds from these stars have been been funnelled outwards in three outflows along 
a NE-SW axis and to the NW, evacuating the region and heating the gas and dust along the path as they are 
swept up in the moving shock front.  

Further, the orientation of the outflow is possibly tilted with respect to the plane of the sky, with the SW 
lobe projecting out of the plane by a few degrees, while the NE lobe is projected into the 
molecular cloud in front of which the cluster sits.  The southern lobe has met with less dense 
material at the border of the molecular cloud and ISM.  This material is being shocked where 
the hot ionised gas of the expanding shell collides with it, as can be seen in Fig.~\ref{fig6}.  
The hot gas can be observed in the  X-ray contours shown in Fig.~\ref{fig6}, filling the expanding cavity 
cleared by the outflow.  The leading shock front is visible at 4.5~${\mu}m$ as shocked Hydrogen emission 
and also via the heated dust visible at 24~${\mu}m$, c.f. Fig.~\ref{fig5} \& \ref{fig6}.   
The expanding flow of hot ionised gas to the NW visible in the X-ray data has caused shocked emission 
in the IR as it collides with and flows around the dust and gas.  
The third outflow, expanding to the NE, appears to be moving into a region of higher dust density and 
appears to be partially obscured by a ridge of dense dust.  
These outflows may have triggered the latest generation of star formation in the central subcluster, 
as can be traced by the filament of protostars,  illustrated in Fig.~\ref{fig6}.   
The YSOs shown in Fig.~\ref{fig6}, identified in $Paper~1$, are the class 0/I (circles) and 
flat spectrum sources (triangles).  While the class 0/I protostars are located throughout the region, they are 
notably concentrated in discreet groups in the three outer subclusters and within the central IRS~2 
subcluster where they form an filament or arc-shaped distribution.  The protostars in this filament are 
also spatially coincident with a ridge of dust emission observed at 1.2mm \citep{vig}.

Outside of the central subcluster, the molecular cloud is also being shaped by other massive stars.   
A dense pillar of dust is being carved away by the photo-evaporating winds of YSO~129 (one of the bow shock 
driving stars) and possibly the O-star candidate ov15-v13, c.f. Fig.~\ref{fig1} and for an expanded view Fig.~\ref{fig2}.   
These two stars lie at projected distances from the top of the pillar of 1.37 and 4.94 arcmin  (0.68 and 2.44~pc) for 
YSO~129 and ov15-v13, respectively.   The pillar has a finger of dense material extending from it in the SW 
direction and two illuminated faces to SW and W.   The faces and finger of the pillar show very bright emission at 
4.5, 8.0 and 24$\mu$m implying both heating of the dust and shocked gas emission.   No point sources 
are detected in the pillar.  This is perhaps because the density is currently too low to promote star formation or 
because any protostars within are as yet too young to be detected in the near- to mid-IR and would require 
observations at longer wavelengths, such as are available with Herschel or with ground-based submm instruments.

Six of the candidate O stars are located on the periphery of the central subcluster.  
If these stars are early type members, then the surrounding low to intermediate mass YSOs that have some shielding 
from the IRS~2 binary would be exposed to higher fluxes of ionising photons than might otherwise have been expected.   
This may have an impact on the X-ray calculated YSO disk fraction (the ratio of X-ray detected class II to class III) 
which was found to be constant with radial distance beyond $\sim$0.5pc from the central IRS~2 binary.  Within a radial 
distance of IRS~2 of 0.5pc the ratio is $40\pm28\%$; this rises to 67$\pm$5\% beyond a radius of 0.5pc and is 
approximately constant from 0.5-3pc from the centre of the subcluster.   While the uncertainties do overlap (due to the 
small sample in the central region), there is an indication that the disk fraction is decreased in the vicinity of IRS~2.  
The three other subclusters  each contain a candidate massive star, and while not as early as IRS~2 they are 
still capable of affecting the circumstellar disks of their subcluster members.  
The majority of the candidates are strewn throughout the distributed population and thus it may be difficult to 
identify any significant fraction of the population of the extended RCW~38 region that is isolated from the effects of 
a massive star.

\subsection{DBS2003-124 \& Protostellar Jet}\label{necj}

The embedded stellar cluster DBS2003-124 was identified by \citet{dut} in a survey of the 2MASS catalogue 
designed to identify new IR clusters.  Located at coordinates $09^{hr}00^m41.5^s$, $-47^d26^m02.3^s$, 16.9' NE 
of  IRS~2,  it lies close to edge of the IRAC image and was only observed in the 3.6 \& 5.8${\mu}m$ bands 
due to the offset in the fields of view of the IRAC channels.  The cluster was also observed in the MIPS 
24${\mu}m$ image, c.f. Fig.~\ref{fig7}.   Visually, the region appears to be contiguous with RCW~38, the cloud 
material and YSOs extending without interruption between this cluster and the subclusters associated with 
RCW~38 itself; therefore we assume the distance to also be 1.7kpc.  At this distance it lies 8.4pc in projection from IRS~2.    

The cluster is surrounded by a ridge of bright emission at 5.8${\mu}m$, within which are located tens of point sources, including 
one very bright 24${\mu}m$ detection in the centre of the grouping and another 24${\mu}m$ source on the NE of 
the central group. A string of five 24${\mu}m$ point sources is observed, two within the ridge of dust emission and
a tail of three more sources extending to the SW, one of which appears to be the origin of the protostellar jet. 
The projected separations of these five objects is semi-regular from the northernmost source to the south: 0.26pc, 0.43pc, 
0.27pc, 0.42pc.   
The southern most of the five 24${\mu}m$ sources, NE20,  is a candidate high mass star, and it is located 1.16~pc 
from the subcluster centre.  This would make DBS2003-124 the only one of the five identified subclusters without 
an O-star at its centre.  Unlike Obj~36, the NW subcluster identified in $Paper~1$, there is no bright mid-IR emission 
nebula associated with this small cluster of stars.  
The region also exhibits faint extended emission at 24${\mu}m$ and is the only subcluster where 24${\mu}m$ point 
sources are identified (though both the central subcluster and Obj~36 are saturated in the 24${\mu}m$ mosaic).

Two methods were used to identify candidate members of the cluster. In the first method, those sources within 
the ridge of dust surrounding the main 24${\mu}m$ source and the 24${\mu}m$ sources themselves were considered 
to be members of the cluster and their coordinates and photometry are included in Table~\ref{tablencphot1}. 
This method identified 14 possible members from the 3.6 and 5.8${\mu}m$ photometry, with the five at 24${\mu}m$. 
The second method involved searching for sources with excess emission on the [3.6] v. [3.6 - 5.8] color-magnitude 
diagram in the region surrounding the cluster and 24${\mu}m$ filament.  With this method, we identify 17 objects, 
including 11 new possible cluster members.  The coordinates and photometry of the sources identified using the 
cmd are presented in Table~\ref{tablencphot2}.  This gives an estimated membership of 33 objects in DBS2003-124.  
High resolution imaging of the cluster in the near-IR would allow for a better estimate of the total population of 
this young cluster.

\subsubsection{The Protostellar Jet}

Figure~\ref{fig7} also reveals two aligned shells of MIR emission, which appear to be shocked emission from 
two infrared candidate HH objects in a westward moving protostellar jet. Tracing the outflow backwards leads to the young star NE21, one of the 
24${\mu}m$ sources in the filament extending from DBS2003-124.  
This star also may have an extended emission feature at 5.8${\mu}m$, which might trace the beginnings of the 
protostellar jet and indicating that the jet remains active, though this feature may be also be a data processing artifact.  
The coordinates of the apexes of the two shells are: $09^{hr}00^m13.94^s$,$-47^d27^m39.05^s$ and $08^{hr}59^m58.93^s$, 
$-47^d27^m54.28^s$. 
These are found at distances from NE21 of 1.85pc and 3.16pc, respectively, assuming a distance of 1.7kpc.  
By considering a range of velocities for the material in the jet, we can estimate the age of the jet from the two identified 
shells. In Table~\ref{jet}, we list 
the ages of the jet assuming velocities of 10, 100 and 500~kms$^{-1}$  \citep{bal2,bal3}.  We estimate the age of the jet to be 
$\sim6\times10^3$ to $3\times10^5$ years.  This would imply a very young age for this star and the subcluster as a whole. 
It is possible that this subcluster is younger than RCW~38 itself and that star formation is progressing through 
the molecular cloud.

\section{\bf Conclusion}

We have examined the RCW~38 cluster using mid-IR observations taken with the {\it Spitzer} Space 
Telescope's IRAC and MIPS instruments and discovered five bow shocks in the  region.  We 
have combined these data with 2MASS photometry and {\it Chandra} observations  to classify the 
evolutionary stage of the driving stars and to estimate their spectral type.  We have examined 
the large scale outflow from the central subcluster to study its effects on the evolution of the cluster as a 
whole, in particular its star formation history.   We have also identified DBS2003-124 as a embedded mid-IR 
subcluster most likely associated with RCW~38 and two candidate HH-objects from a jet originating from a YSO within 
this subcluster.    The main results from our study are:

\begin{itemize}

\item  Five bowshocks were detected by their mid-IR emission in the IRAC and MIPS bandpasses.  
The shocks are associated with driving stars with estimated spectral types from O9-B3, based on 
their near-IR magnitudes and colors.  Four of the driving stars are previously identified YSOs, with 
one class I, two class II and one class III objects.  The fifth driving star was classified as being class III.  
Three of the bow shocks are clearly visible in the IRAC images, two are only clearly detected at 24~${\mu}m$.    

\item  The relative velocities of the shocks were estimated from the stand-off distance of the shocks to their 
associated stars and their estimated mass loss rates and stellar wind velocities.  Though with large 
uncertainties, the values of  $v_{o} n_{o,3}^{1/2}$ were all found to be on the order of 10~km/s.  
Those shocks likely interacting with the ISM, YSO 129 \& 803, and only detected in the longer wavelengths,  
had slightly lower values.   

\item The central O star binary, IRS2, is driving a powerful stellar wind that is shaping the cluster environment. 
The outflow of hot ionised gas, traced by diffuse X-ray emission, is carving a cavity out of the molecular cloud, 
opening the region to the SW and NW where hot ionised gas flows outwards.  
Outflowing material to the north-east  has possibly triggered a new generation of star formation in the region. 

\item A dense pillar of gas and dust has been observed that is being photo-evaporated by the winds from 
the bow-shock driving star YSO~129 and the O-star candidate ov15v13.  No point sources are identified in the 
pillar at near- to mid-IR wavelengths.   

\item We report on the identification in IRAC bands 3.6 \& 5.8${\mu}m$ of DBS2003-124, an IR cluster to the 
NE of RCW~38, as a subcluster of the RCW~38 complex.  Using association with the ridge of bright 5.8${\mu}m$ 
emmision and color-magnitude criteria, we have identified 33 sources as likely members of this cluster. 

\item Two infrared candidate HH-objects from a  protostellar jet, associated with NE21 in DBS2003-124, have been identified.  
Two shocked emission lobes are visible to the W of NE21, at distances of 1.86 and 3.16~pc from NE21.  
Estimating the velocity of the jet material gives an age of  $3\times10^3$ to $3\times10^5$ years for the jet.   

\end{itemize}

This work is based on observations made with the {\it Spitzer} Space Telescope (PID 20127), 
which is operated by the Jet Propulsion Laboratory, California Institute of Technology under NASA 
contract 1407. Support for this work was provided by NASA through contract 1256790 issued by 
JPL/Caltech. Support for the IRAC instrument was provided by NASA through contract 960541 issued 
by JPL.
This publication makes use of data products from the Two Micron All Sky Survey, which is a 
joint project of the University of Massachusetts and the Infrared Processing and Analysis 
Center/California Institute of Technology, funded by the National Aeronautics and Space 
Administration and the National Science Foundation.
This research has made use of the NASA/IPAC Infrared Science Archive, which is operated by 
the Jet Propulsion Laboratory, California Institute of Technology, under contract with the 
National Aeronautics and Space Administration.
This research made use of Montage, funded by the National Aeronautics and Space Administration's 
Earth Science Technology Office, Computation Technologies Project, under Cooperative Agreement 
Number NCC5-626 between NASA and the California Institute of Technology. Montage is maintained 
by the NASA/IPAC Infrared Science Archive.
TLB acknowledges support from NASA through a grant for HST program 11123 from the Space Telescope 
Science Institute, which is operated by the Association of Universities for Research in Astronomy, 
Incorporated, under NASA contract NAS5-26555.

\section{\bf Appendix: Details of Individual Bow Shock sources.}

{\bf YSO 581}   The bow shock associated with YSO 581 is located in the  heart of the RCW~38 cluster 
and likely arises from interaction with the strong winds emanating from the O5.5 binary IRS2. The probable 
driving star (YSO 581) is classified as being of evolutionary class 0/I. It was not detected 
as an X-ray source in the {\it Chandra} observation, and is not a candidate IR variable source.   
The bow shock is observed in all four IRAC bands but due to the high background and source 
density in the central cluster it was difficult to extract accurate values of the flux. It does not have 
a MIPS  detection due to the saturation of the the 24${\mu}m$ mosaic in the central subcluster.  
From the image, the bow shock appears to lie partially in front of the driving star, which would lead to the 
measured fluxes of both objects being contaminated; thus it is possible that the identification of this object 
as having an envelope is inaccurate and that this 'envelope' is in fact emission from the shock.  Likewise the 
 fluxes for the shock are contaminated by the stellar flux, though the aperture used 
to measure the shock flux excluded a circular aperture covering the star itself.  Spectroscopic data of the star 
showing absorption features would help to clarify if this is the case.   
\citet{smi} have identified YSO~581 as source 'D' in their near-IR maps, where they mention the 
possibility of a cavity and ridge of emission to the west of IRS~1 likely corresponding to emission from 
the bow shock.

{\bf YSO 293}  The bow shock associated with YSO 293 was found to the east of IRS~2 in an 
area not immediately affected by the winds from the central massive stars. The young star YSO 293 
was classified as a class II member of the cluster.   It is not a variable candidate, but was detected 
as an X-ray source in the {\it Chandra} observation. The source was detected with 
10 counts in the ACIS data, with $N_H = 1.91\pm0.81\times10^{22}~cm^{-2}$ and $kT = 1.14\pm0.59~keV$.  
The star is detected in the 2MASS bands, and in the three shortest wavelength IRAC bands, but not as a point source 
at 8.0~${\mu}m$.  The bow shock was detected at all four IRAC bands 
and  at 24~${\mu}m$.  The bow shock and star are clearly separated on the sky in the shorter 
IRAC bands, though there is possibly some contamination at 8.0~${\mu}m$ and likely a larger 
degree of contamination at 24~${\mu}m$.   The shock likely arises at the interface of the 
stellar wind with that of the O-star candidate ov15-v13 (c.f. $Paper~1$ Table~12).  This star lies 0.88' to the east, or 0.45pc 
at the assumed distance to RCW~38 of 1.7kpc.   The star is located in a region of elevated emission 
from hot ionised gas in the $Chandra$ data where two candidate O-stars (ov15-v13 \& y244) are also found, 
as shown in Fig.~\ref{fig5}.

{\bf YSO 49}  The next bow shock is associated with YSO 49 and displays similar properties to YSO 293.  
This star is also to the east of IRS~2 and to the south west of YSO 293.  The driving star, YSO 49, was 
classified as being class II. This classification is not affected by the emission from the shock.  
It was not detected in X-rays, though it does lie within the field of view of the observation.  
This star did not show any indication of variability over the two epochs of IRAC observations. The star is 
detected in all 2MASS and IRAC bandpasses, where its stellar SED indicates excess emission in the IR. 
The shock is detected in the four IRAC bands and also in MIPS 24~${\mu}m$.   This star appears to be 
deeply ensconced in the surrounding molecular cloud.   This shock also likely arises at the interface of 
the stellar winds  from ov15-v13 and YSO~49. YSO~49 lies at a distance of 3.4' or 1.68pc from ov15-v13.

\begin{figure}
\epsscale{1.}
\plotone{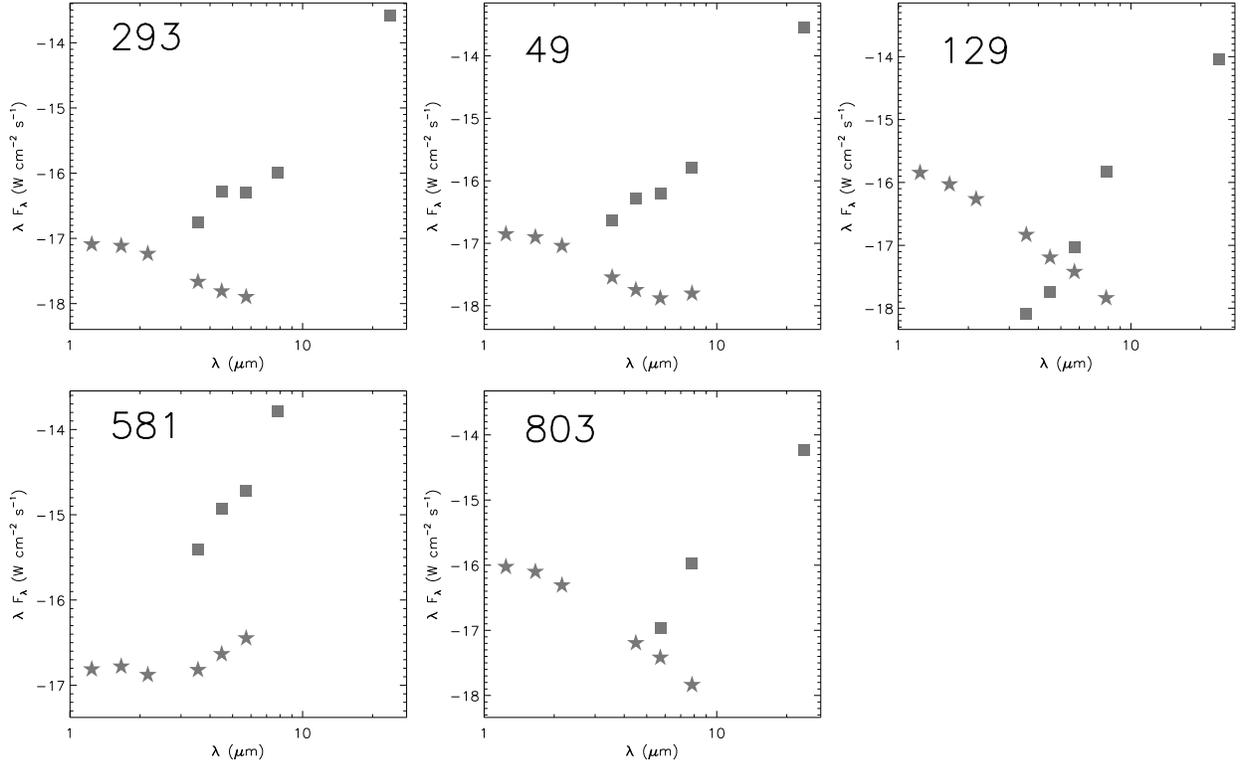}
\caption[]{Spectral Energy Distributions of the five driving stars (stars symbols) and their shocks (square symbols). 
The 24${\mu}m$ fluxes of YSOs 293 and 49 are contaminated by stellar emission. All fluxes for the YSO 581 shock are 
heavily contaminated by variable background and the relative location of 581 behind the shock. 
YSO 129 and source YSO~803 have only weak mid-IR fluxes and are identified at 24${\mu}m$. } 
\label{fig3}
\end{figure}

\begin{figure}
\epsscale{1.}
\plotone{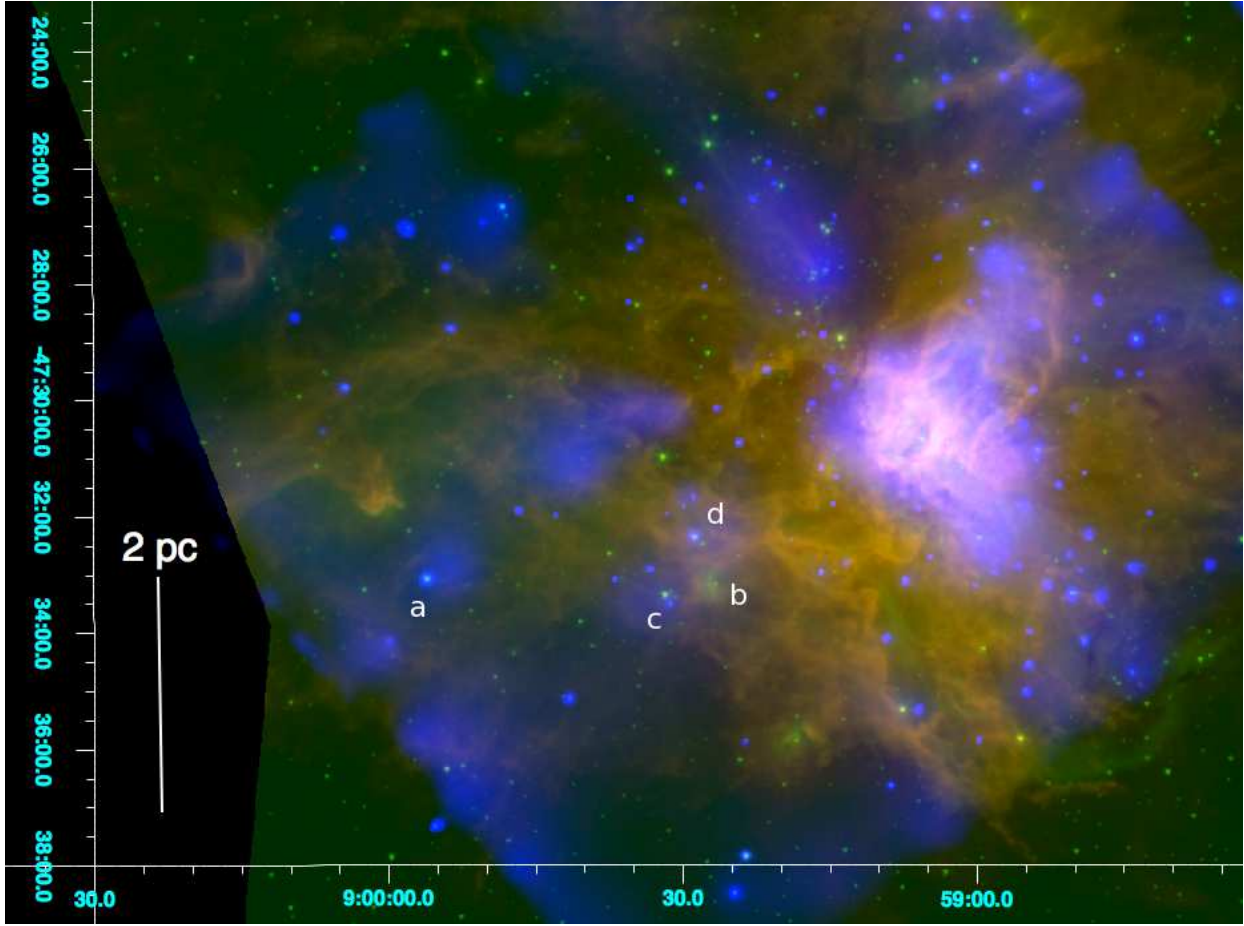}
\caption[]{ Three band false colour image of RCW~38, with the smoothed ACIS-I 0.5-8~keV in blue, IRAC 4.5~${\mu}m$ in green and 8.0~${\mu}m$ in red. 
Hot ionised gas outflows can be seen in the ACIS-I data. In the IRAC bands, ridges of emission from warm dust (8.0~${\mu}m$) 
and shocked hydrogen (4.5~${\mu}m$) coincide with the boundaries of the ionised gas outflows. 
Two of the stars with bow shocks, YSO 129 \& YSO 293 (labelled a \& b, respectively), also lie in regions of ionised gas, driven by winds from 
local massive stars, YSO 129 itself and two candidate O stars identified in $Paper~1$: oc15-v13 (c) and YSO 244 (d).  } 
\label{fig5}
\end{figure}

\begin{figure}
\epsscale{1.}
\plotone{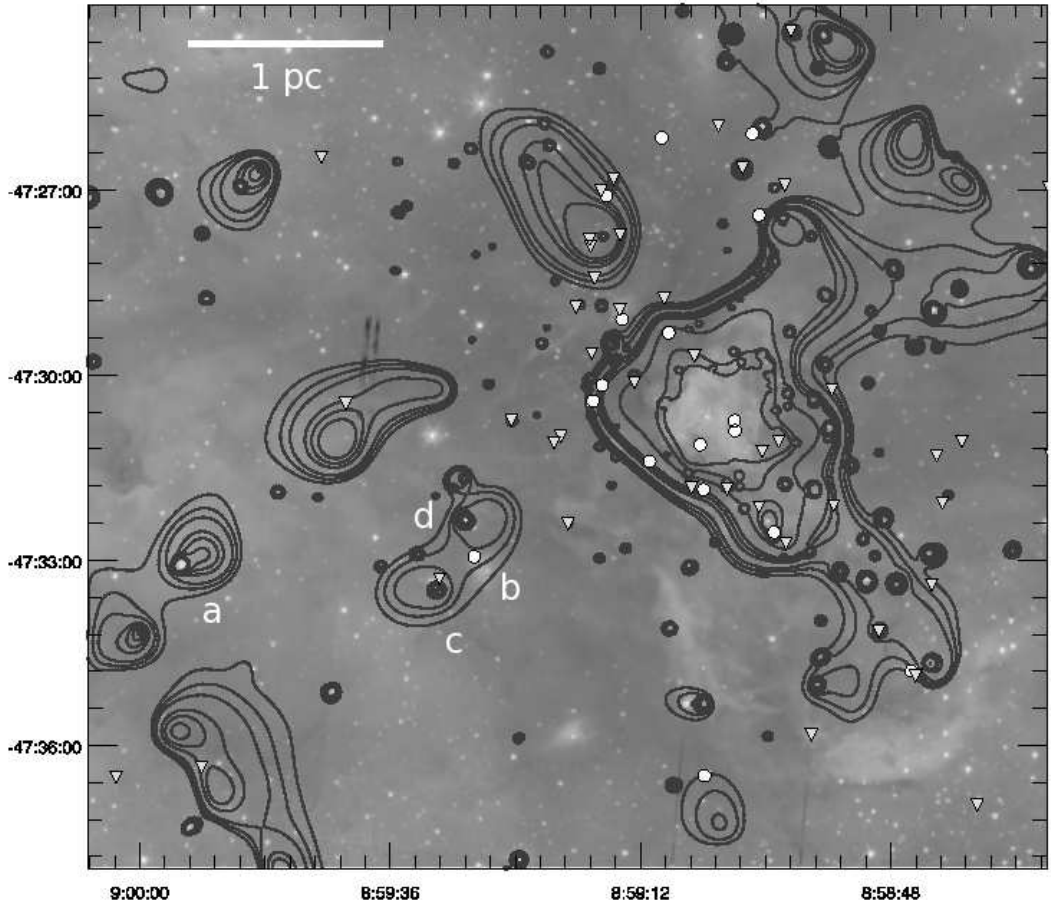}
\caption[]{ Grayscale of the difference between the 4.5 and 5.8${\mu}m$ IRAC image showing the central subcluster with ACIS-I data 
overlaid in contours.   The subtracted image highlights the shocked hydrogen emission in the 4.5${\mu}m$ band in white.  Three outflows 
can be observed, extending to the SW, NE, and NW in the ACIS-I contours. The class 0/I (circles) and flat spectrum (inverted triangles) 
protostars in the central core are overlaid showing the filament of possibly triggered star formation by the NW lobe of the outflow.   
Labels a-d as in Figure~\ref{fig5}.  
 } 
\label{fig6}
\end{figure}

\begin{figure}
\epsscale{0.9}
\plotone{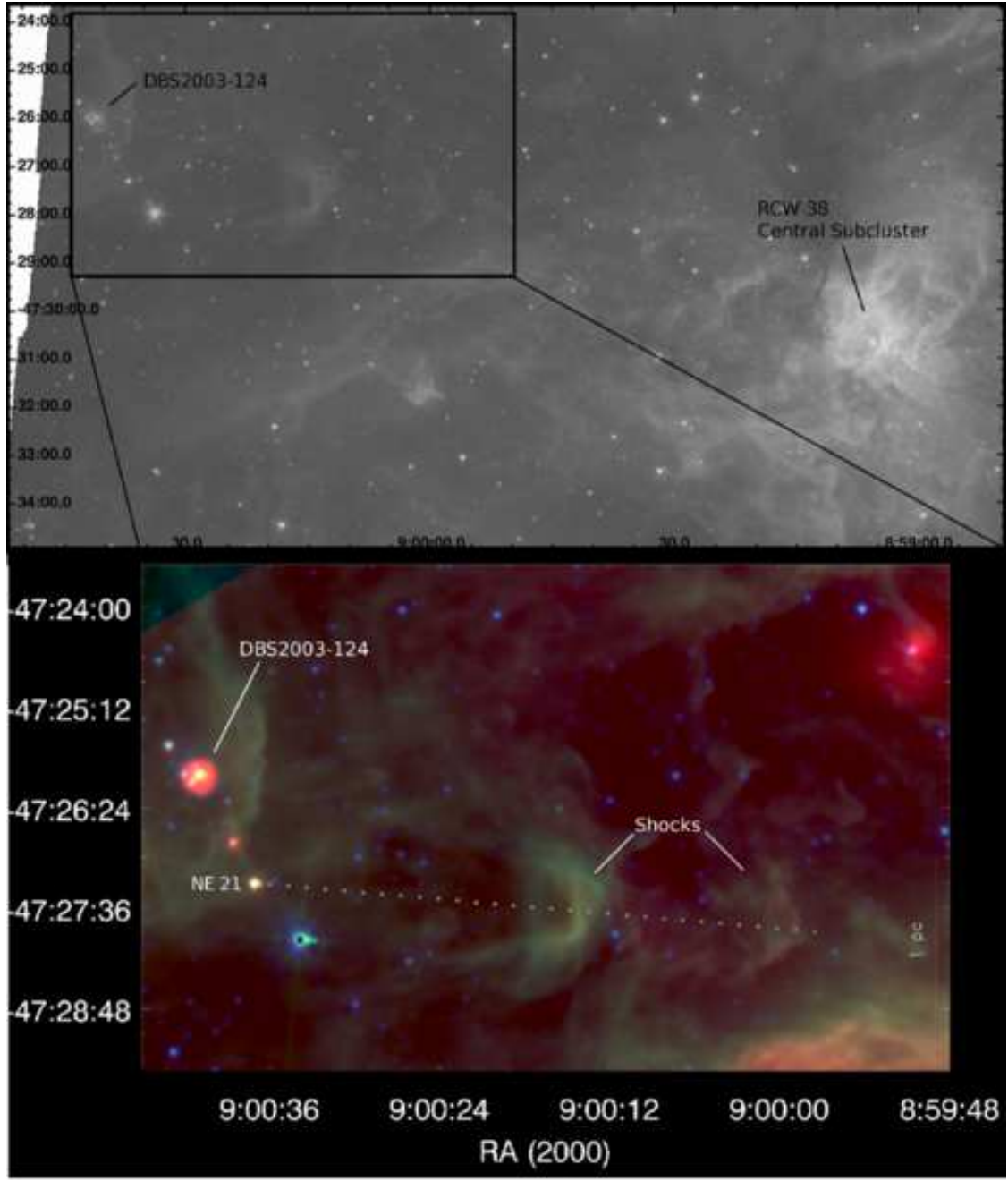}
\caption[]{ {\bf Above:} Grayscale of the 3.6${\mu}m$ IRAC image showing the location of DBS2003-124 with respect to the centre of RCW~38.    
{\bf Below:} Three band false colour image of the newly identified NE subcluster in RCW~38, DBS2003-124, and the YSO-driven jet, 
with IRAC 3.6${\mu}m$ in blue, 5.8${\mu}m$ in green, and MIPS 24${\mu}m$ in red.     The subcluster, in the upper 
left of the image, contains one source with bright 24${\mu}m$ emission, which is surrounded by tens of IRAC sources. 
Two candidate infrared HH-objects associated with the jet are visible in the centre and centre right of the image. The star powering the jet, NE21,  
lies NE of the brightest blue star in the image (OB candidate NE20). 
 } 
\label{fig7}
\end{figure}

{\bf YSO 129}  The fourth bow shock is located further east of IRS~2 than YSOs 293 
and 49.  The driving star is  YSO 129, which has an evolutionary 
classification of class III, indicating that it does not have an optically thick circumstellar disk, consistent with 
its SED in Fig.~\ref{fig3}.  This object was identified as a YSO from its detection in the {\it Chandra} observation.  
The source was detected with 143 counts in the ACIS data, with $N_H = 0.23\pm0.08\times10^{22}~cm^{-2}$ 
and $kT = 1.27\pm0.16~keV$.   It is not a variable candidate.  
The star is detected at all 2MASS and IRAC wavelengths, but not at 
24~${\mu}m$.  The bow shock is visible only at 8.0~${\mu}m$ and 24~${\mu}m$ - indicating that it is 
perhaps more evolved than the others or is located in a less dense medium where the shocked dust 
is not heated to as high a temperature. It may also indicate fine structure emission from a faster shock \citep{neu}.  
Measurements of the fluxes at 3.6, 4.5 and 5.8~${\mu}m$ were also obtained but are comparatively very faint.  
This star's bow shock is possibly interacting with NE20, a new candidate O-star detected in the newly identified NE 
subcluster (c.f. Sec.~\ref{necj}). However, while the projected 2-D orientation of the shock is consistent, 
the large distance, 3.96pc, and the topography of the dust structures surrounding YSO 129 would imply 
that the shock may be due to interactions with the local ISM.  The star is located in a region of diffuse 
X-ray emission, which appears to show extended jet-like emission at the star, see Fig.~\ref{fig6}.  
YSO~129 also illuminates a dense pillar of dust, which it appears to be photo-evaporating, c.f. Sec.~\ref{ow}.

{\bf YSO 803}  The last of the five bow shocks is visible only at 24~${\mu}m$ and is located further to the 
south east of the central RCW~38 cluster than the other shocks, in a region at the edge of the illuminated cloud.  
The driving star was not  identified as a YSO in $Paper~1$;  its SED does not 
show any indication of excess emission in the mid-IR bands indicating that it could be a class III  diskless member. 
It is outside the field of view of the {\it Chandra} observation hence its X-ray flux could not be measured.  
The star YSO~803 is detected in 2MASS and the three shorter IRAC bands, the emission at 8.0~${\mu}m$ 
was extended.  Measurements of the flux of the bow shock were obtained at 5.8, 8.0 and 24~${\mu}m$, 
though the flux at 5.8~${\mu}m$ is very faint.  The detection only at longer wavelengths may suggest that 
this shock has a higher velocity than the other four, as evidenced by fine structure emission at 24~${\mu}m$ 
\citep{neu}.  
This bow shock may also be interacting with the winds from the candidate O-star NE20, though 
the distance of 16' or 8pc means that even a small difference to the orientation of the bow shock will 
miss the star, we therefore conclude that this star is also interacting with the ISM or ambient cluster medium.



\clearpage

\center
\begin{deluxetable}{cccccccccc}
\tablecolumns{10}
\tabletypesize{\scriptsize}
\setlength{\tabcolsep}{0.03in}
\tablewidth{0pt}
\tablecaption{Coordinates and photometry of the stellar sources driving bow shocks. \protect\label{tablestars}}
\tablehead{  \colhead{{\it Sp.}\,\tablenotemark{a}} & \colhead{RA}   &  \colhead{Dec} & \colhead{J} & \colhead{H}  & \colhead{K}   & \colhead{3.6${\mu}m$}  & \colhead{4.5${\mu}m$}  &   \colhead{5.8${\mu}m$}  &  \colhead{8.0${\mu}m$}    \\
\colhead{ID}  &  \colhead{J2000}   &  \colhead{J2000} &  \colhead{[mag.]}  &  \colhead{[mag.]}  &  \colhead{[mag.]}  &  \colhead{[mag.]}   &  \colhead{[mag.]}  &  \colhead{[mag.]}  &  \colhead{[mag.]}    }
\startdata

293  &      8:59:26.70  &  -47:33:10.14   &  11.696$\pm$0.024    &   10.967$\pm$0.027    &    10.503$\pm$0.030    &   10.082$\pm$0.061  &      9.716$\pm$0.101    &    9.168$\pm$0.210   &   \nodata    \\

49    &      8:59:18.44  &  -47:35:51.89     &  11.104$\pm$0.023    &    10.438$\pm$0.026   &     10.016$\pm$0.025     &   9.781$\pm$0.024    &    9.555$\pm$0.035   &    9.125$\pm$0.083   &     7.944$\pm$0.132     \\

129  &     8:59:56.06   &  -47:33:04.4      &    8.587$\pm$0.021    &     8.254$\pm$0.038    &      8.073$\pm$0.021   &      7.999$\pm$0.009   &       8.169$\pm$0.012   &      7.978$\pm$0.014     &     8.029$\pm$0.031 \\

581  &     8:59:02.94   &  -47:30:54.1      &    11.002$\pm$0.021    &    10.130$\pm$0.021    &     9.607$\pm$0.019    &     7.966$\pm$0.038     &     6.772$\pm$0.032      &      5.544$\pm$0.072   &   \nodata      \\

803  &     8:59:41.98   &  -47:41:22.8      &    9.035$\pm$0.025    &    8.434$\pm$0.042     &     8.186$\pm$0.030    &      8.879$\pm$0.015   &     8.175$\pm$0.012     &    7.970$\pm$0.014       &     8.027$\pm$0.021    \\
             
\enddata
\tablenotetext{a}{See \citet{winR} ($Paper~1$) for further details. \\  No photometry was available for the sources at 24${\mu}m$.} 

\end{deluxetable}
\endcenter

\center
\begin{deluxetable}{cccccccc}
\tablecolumns{8}
\tabletypesize{\scriptsize}
\tablewidth{0pt}
\tablecaption{Coordinates and photometry for the bow shocks. \protect\label{tableshocks}}
\tablehead{  \colhead{{\it Sp.}} & \colhead{RA$_{apex}$}   &  \colhead{Dec$_{apex}$} &  \colhead{3.6${\mu}m$}  & \colhead{4.5${\mu}m$}  &   \colhead{5.8${\mu}m$}  &  \colhead{8.0${\mu}m$}   &  \colhead{24${\mu}m$}  \\
\colhead{ID}  &  \colhead{J2000}   &  \colhead{J2000}   &  \colhead{[mJy]}   &  \colhead{[mJy]}  &  \colhead{[mJy]}  &  \colhead{[mJy]} &  \colhead{[mJy]}   }
\startdata

293\_s    &   8:59:27.31   &    -47:33:09.0   &   16.3$\pm$1.8  &      38.2$\pm$2.1    &    28.9$\pm$3.8   &  43.1$\pm$6.2    &   $<$3598    \\

49\_s    &    8:59:18.62   &   -47:35:45.8     &   21.6$\pm$1.3    &    38.9$\pm$1.6   &   36.3$\pm$2.7   &     68.6$\pm$4.8  &   $<$4013    \\

129\_s  &   08:59:57.36  &   -47:32:45.7    &   $\le$0.8     &    $\le$1.4    &    5.4$\pm$5.0    &   62.7$\pm$25.4     &   1251$\pm$14    \\  

581\_s  &   08:59:02.94  &   -47:30:54.1    &   $\le$366    &    $\le$869    &  $\le$1127      &   $\le$6950    &   \nodata    \\  

803\_s  &   08:59:43.07  &   -47:41:28.6    &   \nodata     &    \nodata    &    6.3$\pm$2.6    &   45.8$\pm$3.0     &   813$\pm$16    \\  
         
\enddata
\end{deluxetable}
\endcenter

\center
\begin{deluxetable}{cccc}
\tablecolumns{4}
\tabletypesize{\scriptsize}
\tablewidth{0pt}
\tablecaption{Stand-off distances and estimated stellar wind properties. \protect\label{tableprops}}
\tablehead{  \colhead{{\it Sp.}} & \colhead{Est. Spectral Type}    &  \colhead{$d_{w}cos i$}  & \colhead{$v_{0}n^{1/2}_{0,3}(cos i)^{-1}$\,\tablenotemark{a}}  \\
\colhead{ID}  &  \colhead{}      &  \colhead{pc}  &  \colhead{$km s^{-1}$}  }
\startdata

293    &   B2-B3     &   0.079  &      $<$8.41   \\

49    &    B1.5-B2      &   0.055    &    $<$12.19    \\
         
129   &   B0        &   0.177   &   $<$3.79    \\         

581   &   B0.5-B1      &   \nodata   &   \nodata   \\         

803   &   B0       &   0.125   &   $<$5.37    \\         
         
\enddata
\tablenotetext{a}{Calculated for an assumed mass loss rate of $<$0.2$\dot{M}_{w,-6}$ for each star.}
\end{deluxetable}
\endcenter

\center
\begin{deluxetable}{cccccccc}
\tablecolumns{8}
\tabletypesize{\scriptsize}
\tablewidth{0pt}
\tablecaption{Coordinates and photometry for the Candidate Members of NE subcluster. \protect\label{tablencphot1}}
\tablehead{  \colhead{{\it Sp.}} & \colhead{RA}   &  \colhead{Dec} &  \colhead{J}   &  \colhead{H}   &  \colhead{K$_s$}   &  \colhead{3.6${\mu}m$}   &   \colhead{5.8${\mu}m$}    \\
\colhead{ID}  &  \colhead{J2000}   &  \colhead{J2000}   &  \colhead{[mag.]} &  \colhead{[mag.]}  &  \colhead{[mag.]}  &  \colhead{[mag.]}  &  \colhead{[mag.]}   }
\startdata

1  &  9:00:42.92  &  -47:25:52.30  &  \nodata  &  15.22$\pm$0.11  &  13.98$\pm$0.08  &  12.48$\pm$0.09  &  11.24$\pm$0.11    \\

2  &  9:00:42.04  &  -47:26:22.32  &  16.06$\pm$0.10  &  14.55$\pm$0.05  &  13.84$\pm$0.05  &  13.16$\pm$0.11  &  12.91$\pm$0.39    \\

3  &  9:00:43.04  &  -47:26:07.38  &  \nodata  &  16.14$\pm$0.19  &  14.81$\pm$0.14  &  13.01$\pm$0.09  &  11.85$\pm$0.18    \\

4  &  9:00:41.42  &  -47:26:05.00  &  16.38$\pm$0.13  &  13.59$\pm$0.06  &  11.75$\pm$0.04  &  9.78$\pm$0.04  &  \nodata    \\

5  &  9:00:39.22  &  -47:26:34.58  &  \nodata  &  14.65$\pm$0.06  &  12.78$\pm$0.03  &  11.54$\pm$0.05  &  11.02$\pm$0.12    \\

6  &  9:00:43.09  &  -47:25:39.47  &  13.82$\pm$0.03  &  12.12$\pm$0.03  &  11.00$\pm$0.02  &  9.37$\pm$0.02  &  7.79$\pm$0.01    \\

7  &  9:00:42.09  &  -47:25:53.74  &  \nodata  &  \nodata  &  13.84$\pm$0.19  &  11.99$\pm$0.06  &  10.55$\pm$0.08    \\

8  &  9:00:41.24  &  -47:25:54.26  &  \nodata  &  \nodata  &  13.11$\pm$0.15  &  11.04$\pm$0.12  &  9.08$\pm$0.04    \\

9  &  9:00:42.01  &  -47:26:00.25  &  \nodata  &  \nodata  &  11.43$\pm$0.05  &  9.98$\pm$0.05  &  8.72$\pm$0.05    \\

10  &  9:00:39.62  &  -47:26:20.94  &  \nodata  &  \nodata  &  \nodata  &  14.82$\pm$0.39  &  \nodata    \\

11  &  9:00:40.88  &  -47:26:14.79  &  \nodata  &  \nodata  &  \nodata  &  12.00$\pm$0.13  &  \nodata    \\

12  &  9:00:40.86  &  -47:25:59.36  &  \nodata  &  \nodata  &  \nodata  &  10.29$\pm$0.07  &  \nodata    \\

13  &  9:00:41.43  &  -47:25:45.97  &  \nodata  &  \nodata  &  \nodata  &  13.29$\pm$0.15  &  \nodata    \\

14  &  9:00:39.29  &  -47:25:42.83  &  \nodata  &  \nodata  &  \nodata  &  14.76$\pm$0.26  &  \nodata    \\

15  &  9:00:41.94  &  -47:25:33.31  &  \nodata  &  \nodata  &  \nodata  &  14.19$\pm$0.21  &  \nodata    \\

16  &  9:00:39.90  &  -47:25:35.15  &  \nodata  &  \nodata  &  \nodata  &  14.90$\pm$0.38  &  \nodata    \\

17  &  9:00:39.31  &  -47:25:36.05  &  \nodata  &  \nodata  &  \nodata  &  14.85$\pm$0.26  &  \nodata    \\

18  &  9:00:40.77  &  -47:26:14.78  &  \nodata  &  \nodata  &  \nodata  &  \nodata  &  10.47$\pm$0.12    \\

19  &  9:00:41.42  &  -47:26:06.34  &  \nodata  &  \nodata  &  \nodata  &  \nodata  &  8.00$\pm$0.04    \\

20  &  9:00:33.80  &  -47:27:58.17  &  8.40$\pm$0.02  &   6.60$\pm$0.03  &  5.76$\pm$0.02  & \nodata &  6.28$\pm$0.01 \\

21  &  9:00:36.96  &  -47:27:19.10  &  \nodata  &  15.77$\pm$0.15  &  12.99$\pm$0.03  &  8.93$\pm$0.01  & 6.44$\pm$0.01  \\
 
22  &  9:00:38.62  &  -47:26:48.98  &  \nodata  &  15.64$\pm$0.13  &  13.51$\pm$0.07  &  11.04$\pm$0.04  &  8.91$\pm$0.03   \\
    
\enddata

\end{deluxetable}
\endcenter

\center
\begin{deluxetable}{cccccccc}
\tablecolumns{8}
\tabletypesize{\scriptsize}
\tablewidth{0pt}
\tablecaption{Coordinates and photometry for the Candidate Members of NE subcluster from 3.6${\mu}m$ v 3.6${\mu}m$-5.8${\mu}m$ Color-Color diagram. \protect\label{tablencphot2}}
\tablehead{  \colhead{{\it Sp.}} & \colhead{RA}   &  \colhead{Dec} &  \colhead{J}   &  \colhead{H}   &  \colhead{K$_s$}   &  \colhead{3.6${\mu}m$}   &   \colhead{5.8${\mu}m$}    \\
\colhead{ID}  &  \colhead{J2000}   &  \colhead{J2000}   &  \colhead{[mag.]} &  \colhead{[mag.]}  &  \colhead{[mag.]}  &  \colhead{[mag.]}  &  \colhead{[mag.]}   }
\startdata

22  &  9:00:38.62  &  -47:26:48.98  &  \nodata  &  15.64$\pm$0.13  &  13.51$\pm$0.07  &  11.04$\pm$0.04  &  8.91$\pm$0.03   \\

23  &  9:00:26.47  &  -47:26:27.64  &  16.05$\pm$0.08  &  14.52$\pm$0.05  &  13.93$\pm$0.05  &  13.46$\pm$0.12  &  12.74$\pm$0.23    \\

24  &  9:00:26.76  &  -47:26:57.52  &  \nodata  &  \nodata  &  \nodata  &  11.91$\pm$0.06  &  11.05$\pm$0.06    \\

25  &  9:00:29.65  &  -47:27:31.30  &  13.89$\pm$0.03  &  13.19$\pm$0.03  &  12.90$\pm$0.04  &  12.79$\pm$0.08  &  12.19$\pm$0.16    \\

1  &  9:00:42.92  &  -47:25:52.30  &  \nodata  &  15.22$\pm$0.11  &  13.98$\pm$0.08  &  12.48$\pm$0.09  &  11.24$\pm$0.11    \\

26  &  9:00:44.90  &  -47:25:20.41  &  \nodata  &  \nodata  &  15.08$\pm$0.15  &  12.56$\pm$0.08  &  11.08$\pm$0.07    \\

3  &  9:00:43.04  &  -47:26:07.38  &  \nodata  &  16.14$\pm$0.19  &  14.81$\pm$0.14  &  13.01$\pm$0.09  &  11.85$\pm$0.18    \\

27  &  9:00:34.01  &  -47:25:53.45  &  \nodata  &  15.07$\pm$0.08  &  14.28$\pm$0.08  &  13.12$\pm$0.10  &  12.20$\pm$0.18    \\

28  &  9:00:36.25  &  -47:25:52.88  &  16.11$\pm$0.10  &  14.46$\pm$0.05  &  13.75$\pm$0.06  &  12.57$\pm$0.07  &  10.80$\pm$0.06    \\

29  &  9:00:38.98  &  -47:26:19.35  &  \nodata  &  \nodata  &  14.82$\pm$0.13  &  11.90$\pm$0.06  &  10.26$\pm$0.06    \\

30  &  9:00:43.83  &  -47:24:04.82  &  \nodata  &  15.87$\pm$0.16  &  14.11$\pm$0.06  &  12.47$\pm$0.07  &  11.30$\pm$0.10    \\

31  &  9:00:35.82  &  -47:24:46.60  &  15.37$\pm$0.06  &  13.82$\pm$0.04  &  12.98$\pm$0.03  &  12.04$\pm$0.06  &  11.14$\pm$0.08    \\

6  &  9:00:43.09  &  -47:25:39.47  &  13.82$\pm$0.03  &  12.12$\pm$0.03  &  11.00$\pm$0.02  &  9.37$\pm$0.02  &  7.79$\pm$0.01    \\

32  &  9:00:43.79  &  -47:24:38.94  &  15.22$\pm$0.05  &  12.99$\pm$0.03  &  11.77$\pm$0.03  &  10.43$\pm$0.03  &  9.32$\pm$0.03    \\

7  &  9:00:42.09  &  -47:25:53.74  &  \nodata  &  \nodata  &  13.84$\pm$0.19  &  11.99$\pm$0.06  &  10.55$\pm$0.08    \\

9  &  9:00:42.01  &  -47:26:00.25  &  \nodata  &  \nodata  &  11.43$\pm$0.05  &  9.98$\pm$0.05  &  8.72$\pm$0.05    \\

33    &  9:00:26.97  &  -47:25:08.63  &  \nodata  &  \nodata  &  \nodata  &  13.58$\pm$0.12  &  12.44$\pm$0.17    \\

\enddata

\end{deluxetable}
\endcenter

\center
\begin{deluxetable}{cccc}
\tablecolumns{4}
\tabletypesize{\scriptsize}
\tablewidth{0pt}
\tablecaption{Jet velocities and ages for the two shells. \protect\label{jet}}
\tablehead{  \colhead{velocity} & \colhead{velocity}   &  \colhead{1st Shell} &  \colhead{2nd Shell} \\
\colhead{$kms^{-1}$}  &  \colhead{$AU yr^{-1}$}   &  \colhead{yr}   &  \colhead{yr}   }
\startdata

10    &  0.21   &   $1.8\times10^5$   &  $3.1\times10^5$   \\

         
100  &   21   &       $1.8\times10^4$   &  $3.1\times10^4$       \\         

500  &   105   &    $3.6\times10^3$   &    $6.2\times10^3$     \\

\enddata

\end{deluxetable}
\endcenter


\clearpage

\clearpage

\clearpage

\clearpage

\clearpage

\clearpage

\clearpage

\end{document}